\documentclass[twocolumn]{aastex62}
\usepackage{apjfonts}
\usepackage{amsmath}

\newcommand{\Hi}{\hbox{H{\sc i}}}

\newcommand{\halpha}{H$\alpha$}
\newcommand{\hbeta}{H$\beta$}
\newcommand{\oiiifull}{$\text{[O\,{\sc iii}]}\lambda \, 5007$}

\newcommand{\niifull}{$\text{[N\,{\sc ii}]}\lambda \, 6583$}

\newcommand{\siifull}{$\text{[S\,{\sc ii}]}\lambda\lambda \, 6716,6731$}

\begin{document}

\title{On the formation of star-forming galaxies having anomalously low-metallicity regions: the role of galaxy interaction and CGM/IGM accretion}
\author{Yibo Cao}
\affiliation{Purple Mountain Observatory, Chinese Academy of Sciences, 10 Yuan Hua Road, Nanjing, Jiangsu 210033, China}
\affiliation{School of Astronomy and Space Sciences, University of Science and Technology of China, Hefei, 230026, China
}

\author{Zhizheng Pan}
\correspondingauthor{Zhizheng Pan}
\email{panzz@pmo.ac.cn}
\affiliation{Purple Mountain Observatory, Chinese Academy of Sciences, 10 Yuan Hua Road, Nanjing, Jiangsu 210033, China}
\affiliation{School of Astronomy and Space Sciences, University of Science and Technology of China, Hefei, 230026, China
}

\begin{abstract}
Gas accretion from both the circum-galactic medium (CGM)/inter-galatic medium (IGM) and interacting companion galaxy can dilute the gas phase metallicity of a galaxy. However, their relative contribution to the chemical evolution of galaxies remains to be quantified. To this end, in this work we study a sample of 510 star-forming galaxies (SFGs) having anomalously low-metallicity (ALM) regions selected from the MaNGA data available in the Data Release 17 from the Sloan Digital Sky Survey. ALM regions are defined as those having gas phase metallicities that are at least $\sim 2\sigma$ lower than the emprical relation between stellar mass surface density ($\Sigma_{*}$) and gas phase metallicity, i.e., the $\Sigma_{*}-Z$ relation. We find that ALM galaxies have higher star formation rates and \Hi~gas fractions than normal SFGs at fixed $M_*$. $\sim$25\% of the ALM galaxies exhibit tidal features, while the tidal fraction is only $\sim$12\% for normal SFGs, indicating that galaxy interaction is an important factor responsible for the ALM phenomenon. To explore the origin of non-tidal ALM galaxies, we compare their morphologies and environments with those of mass-matched normal SFGs. We find that non-tidal ALM galaxies tend to have more disk-dominated morphologies and reside in less-dense environment. These findings suggest that cold gas accretion from the CGM/IGM is the primary cause for the ALM phenomenon, while galaxy interaction plays a minor but non-negligible role.
\end{abstract}


\section{Introduction} \label{sec:intro}
Gas accretion (or gas inflow) has been widely acknowledged as the key driver of galaxy growth, providing cold gas as the fuel for star formation that leads to the build-up of stellar mass of a galaxy \citep{Dekel 2009, Papovich 2011,Somerville 2015,Rodri 2016}. This process can be achieved either by accreting gas from the surrounding medium, such as the circum-galactic medium (CGM)/inter-galactic medium (IGM) \citep{McQuinn 2016, Tumlinson 2017, Faucher 2023}, or by stripping gas from companion galaxies through galaxy interactions. However, direct observation of on-going gas accretion events remains quite challenging for extragalactic galaxies \citep{Steidel 2010, Bouche 2013,Ju 2022}.

Recently, the gas phase metallicity has been used as an indirect tracer to probe gas inflows of galaxies \citep{Hwang 2019, Luo 2021,Pace 2021,Delvalle 2023,Olvera 2024}. This is because the inflow gas typically has lower metallity than the pre-existing gas in the galaxy. When the two kinds of gas getting mixed, the dilution effect occurs in the gas phase metallicity \citep{Luo 2021}.  Using the MaNGA data, \citet{Hwang 2019} have shown that a significant fraction of disk SFGs ($\sim 25$\%) have regions within which the gas phase metallicity (12+log(O/H)) is anomalously low compared to expectations from the empirical relation between stellar mass surface density and metallicity. Interestingly, these anomalously low-metallicity (ALM) regions tend to have high specific star formation rate (sSFR), and are preferentially found in the outer regions of low-mass galaxies. In addition, the incidence rate of ALM regions is high in morphological disturbed galaxies. The authors thus argue that the ALM regions are the sites in which the low metallicity gas has been recently accreted and trigger star formation. \citet{Luo 2021} study the N/O abundance ratios of ALM regions and show that ALM regions typically have higher N/O abundance ratio at a given O/H in the N/O versus O/H plane. This finding nicely agrees with the scenario that ALM regions are produced by the mixing of pre-existing metal-rich gas and recently accreted metal-poor gas. In addition, it rules out the alternative interpretation that ALM regions are just less chemically evolved, in which case they will follow a similar distribution as normal regions in the the N/O versus O/H plane.

Given the important role of gas accretion in galaxy evolution, SFGs hosting ALM regions are of great interest and need to be studied in more detail. \citet{Hwang 2019} have selected 307 out of 1222 late-type SFGs as their ALM galaxy sample and studied their properties, such as morphologies and star formation rates (SFRs). However, a number of questions remain. For example, what is the relative role of galaxy interaction and CGM/IGM accretion to the formation of ALM galaxies? Does the formation of ALM galaxies depend on environments? In addition, \citet{Hwang 2019} argued that ALM galaxies have experienced gas accretion recently, but the authors do not investigate the cold gas content of ALM galaxies. To answer these questions, complementary studies based on a larger sample of ALM galaxies are needed.

With the complete MaNGA galaxy sample released in SDSS DR17, it is now possible to select a significantly larger sample of ALM galaxies than \citet{Hwang 2019} and perform an in-depth analysis to answer the questions listed above. This paper is organized as follows. Section~\ref{sec:data} describes the data used for analysis and sample selection. Section~\ref{sec:results} presents the results. Our discussion is briefly stated in Section~\ref{sec:discussion} and the conclusion is summarized in Section~\ref{sec:conclusion}. In this work,we adopt a concordance $\Lambda$CDM cosmology with $\Omega_{\rm m}=0.3$, $\Omega_{\rm \Lambda}=0.7$, $H_{\rm 0}=70$ $\rm km~s^{-1}$ Mpc$^{-1}$ and a \citet{Chabrier 2003} initial mass function (IMF).

\section{Data} \label{sec:data}

MaNGA is an integrated field spectroscopy (IFS) survey carried out in the SDSS-IV \citep{Gunn 2006, Bundy 2015,Blanton 2017}, aiming to provide spatially resolved spectroscopic information for $\sim$ 10,000 galaxies at $z<0.15$. The sizes of integrated field units (IFUs) used in MaNGA vary from 19 to 127 fibers, which are designed to cover the target galaxies out to at least 1.5~$R_{\rm e}$, where $R_{\rm e}$ is the half-light radius. The wavelength coverage of MaNGA is from 3600 to 10300 \AA, with a resolution of $R \sim 2000$ and a reconstructed point spread function (PSF) of $\sim$ 2.5 arcsec in full width at half maxima (FWHM) \citep{Bundy 2015, Law 2016}. The MaNGA survey has been completed and the data products are publicly released in SDSS data release 17 (DR17) \footnote{https://www.sdss4.org/dr17/manga/}. In this study, the galaxy sample is exacted from the MaNGA Pipe3D value-added catalog \citep{Sanchez 2022}.

For the sake of studying the physical properties of MaNGA galaxies, we further include data from the value-added catalogs (VACs) of SDSS. The \Hi~data are drawn from the \Hi-MaNGA catalog \citep{Masters 2019, Stark 2021}, which contains the \Hi~information of 3358 targets taken from the Green Bank Telescope (GBT) observations, as well as $\sim 3300$ targets from the Arecibo Legacy Fast ALFA (ALFALFA) survey \citep{Haynes 2018}. The morphological classifications are from the MaNGA Visual Morphology catalog, which contains visual morphological classifications based on inspections of image mosaics using a new re-processing of SDSS and Dark Energy Spectroscopy Instrument (DESI) Legacy Survey images \citep{Dey 2019, Vaz 2022}. In addition to the Hubble type classifications, labels of "Tidal galaxies" are also provided in this catalog. To investigate the environments of our sample galaxies, the large-scale structure (LSS) information is from the Galaxy Environment for MaNGA (GEMA) catalog \citep{Arg 2015, Ether 2015, Wang 2016}. In addition, the local galaxy environment estimates are from the Cosmic Slime catalog, in which the local matter density at the locations of galaxies are derived using the Monte Carlo Physarum Machine (MCPM) algorithm inspired by the growth and movement of Physarum polycephalum slime mold \citep{Burch 2020,Wild 2023}. It is worth noting that not every individual MaNGA galaxy is included in these VACs. Given this, in the following sections we also specify the number of galaxies when cross-matching our sample galaxies with the used VAC.

\subsection{Sample Selection} \label{subsec:sample}

Following a similar procedure presented in \citet{Hwang 2019}, we first select disk galaxies based on the Sersic index $n$, with $n<2.5$ \citep{Sersic 1963,Shen 2003, Lange 2015}. To exclude edge-on objects, only galaxies with minor-to-major axis ratio of $b/a>0.35$ are selected. The stellar mass range is then limited between $10^{8}M_{\odot}$ and $10^{12}M_{\odot}$. For individual galaxies, the Pipe3D datacubes provide a mask map to mask out foreground stars and spaxels whose signal-to-noise ratio (S/N) of emission line measurement is unreliable. So we exclude bad spaxels with the Pipe3D mask map. As done by \citet{Hwang 2019}, we only consider spaxels with deprojected stellar surface mass density $\Sigma_{*}> 10^{7}\, M_{\odot}\, \mathrm{kpc^{-2}}$.  All the physical parameters of galaxies are extracted from the Pipe3D catalog.

\begin{figure}
\centering
\includegraphics[width=80mm,angle=0]{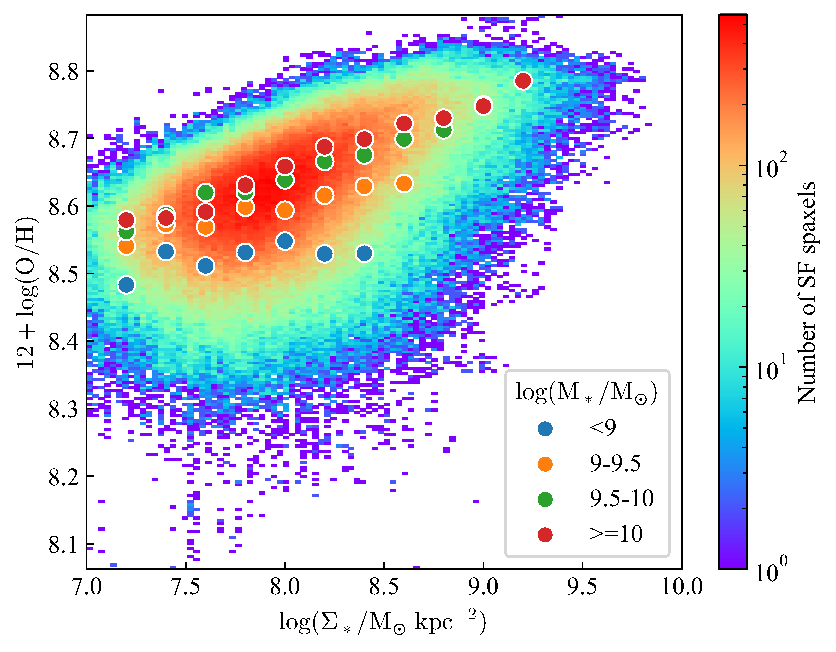}
\caption{The relation between metallicity and stellar surface mass density, i.e., the local $\Sigma_{*}-Z$ relation. The circles represent the peaks of metallicity distributions in each stellar mass and surface mass density bin. The colorful background shows the distribution of all pure star-forming spaxels in late-type galaxies with stellar mass $> 10^8 M_{\odot}$. The bin of $< 10^9 M_{\odot}$ is only used for interpolation, and we do not further analyze the galaxies in this bin.}\label{fig1}
\end{figure}

We use the BPT diagram to select star-forming (SF) spaxels \citep{BPT 1981}. Specifically, SF spaxels are selected using the \citet{Kauffmann 2003} demarcation curve. Meanwhile, we demand the equivalent width of \halpha~(EW(\halpha)) to be greater than 3 \AA~to exclude emissions produced by potentially low ionization sources \citep{Cid 2011,Belfiore 2016}. Moreover, the S/Ns of the emission lines used in the BPT diagram and later metallicity calculations, including \halpha, \hbeta, \oiiifull, \niifull, \siifull, are limited to be S/N$>10$. Only SF spaxels are included in our following analysis. Finally, a galaxy is classified as a star-forming galaxy (SFG) if its number of SF spaxels is $>=20$, considering that the angular area of a typical MaNGA PSF is 19.6 spaxels.\footnote{The FWHM of SDSS PSF is 2.5 arcsec, so the angular area of PSF is $S_{\rm psf}=\pi (\rm FWHM/2)^2=3.14(2.5/2 ~arcsec)^2=4.90 ~arcsec^2$. The angular area of a MaNGA pixel is $S_{\rm pix}=0.25~\rm arcsec^2$. So the angular area of MaNGA PSF corresponds to $N=S_{\rm psf}/S_{\rm pix}=4.9/0.25=19.6$ spaxels.}

After applying these selection criteria, we obtain a sample of 2369 disk SFGs out of the 10,220 MaNGA galaxies, and these SFGs contain 599,313 SF spaxels. In the following analysis, we will focus on the 2128 disk SFGs with $M_{*}>10^{9}M_{\odot}$.

\subsubsection{Definition of Anomalously Low-metallicity (ALM) Galaxy} \label{subsubsec:alm}

In this work, we use a strong line calibrator:  RS32 =\oiiifull /\hbeta~+ \siifull /\halpha~from \citet{Curti 2020} to calculate gas phase metallicity. In order to check whether the results are dependent on the choice of metallicity calibrator, we also use the O3N2 calibrator of \citet{Curti 2020} to repeat the analysis (see Appendix). We show that the results are unchanged, suggesting that our findings do not depend on the used metallicity calibrator.

\begin{figure}
\centering
\includegraphics[width=80mm,angle=0]{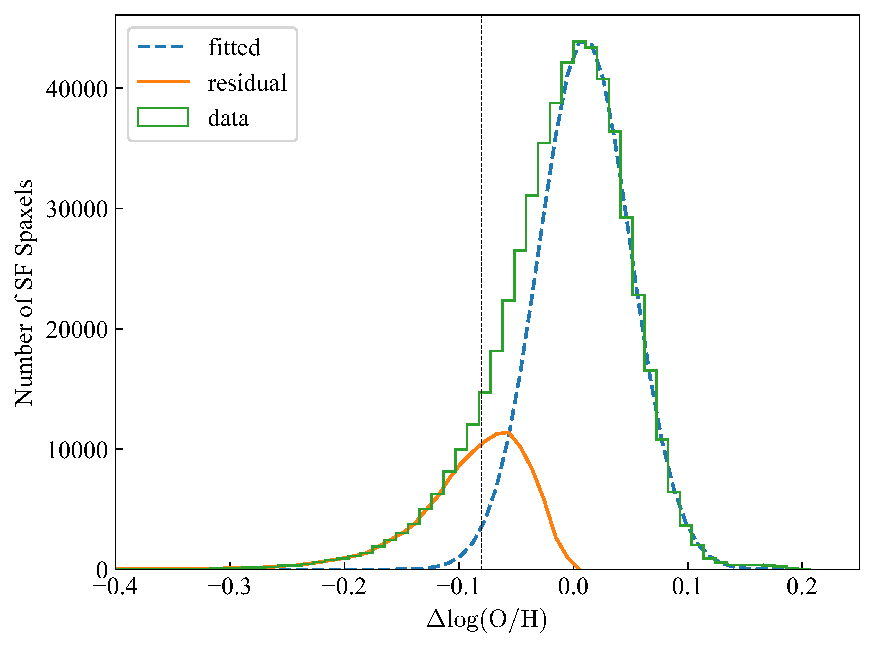}
\caption{Distribution of the metallicity deviation of SF spaxels in the late-type galaxies with stellar mass $>10^9 M_{\odot}$ (green histogram). The blue dashed curve is a Gaussian fitted profile for "normal" spaxels, with mean value at 0 and a standard deviation of $\sigma_{\rm Z} = 0.04$. The orange solid curve is the residual of the fitted Gaussian fitted profile subtracted from the negative side of the histogram. The vertical black dotted line marks where $\Delta\log(\mathrm(O/H)) = -0.08$. ALM spaxels are selected as those with $\Delta\log(\mathrm(O/H)) < -0.08$.}\label{fig2}
\end{figure}

The ALM regions are selected following the same procedure as presented in \citet{Hwang 2019} and \citet{Luo 2021}. First, the expected metallicity $\mathrm{(12+\log(O/H))_{exp}}$ for SF spaxels are derived via the interpolation of the empirical relation between stellar surface mass density and gas phase metallicity at fixed stellar mass, i.e., the local $\Sigma_{*}-Z$ relation. The metallicity deviation, defined as $\Delta\mathrm{\log(O/H) = (12 + \log(O/H))_{obs} - (12 + \log(O/H))_{exp}}$, is then used to select spaxels with anomalously low metallcity. Figure~\ref{fig1} shows the local $\Sigma_{*}-Z$ relations for different mass bins.

Figure~\ref{fig2} illustrates the distribution of metallicity deviation $\Delta\mathrm{\log(O/H)}$. The distribution of all the SF spaxels are shown in the green histogram, which has an obvious low-metallicity tail. To characterize the "normal" spaxels, we fit the positive side of the observed $\Delta\mathrm{\log(O/H)}$ distribution with a Gaussian profile. The best fit model is shown in the dashed blue line, with a dispersion of $\sigma_{Z}=0.04$. This is in good agreement with the result of \citet{Luo 2021}, in which $\sigma_{Z}=0.037$. In \citet{Hwang 2019}, the authors find a dispersion of $\sigma_{Z}=0.07$. This difference is likely originated from the different metallicity calibrators used in these two studies. We define spaxels with $\Delta\log(\mathrm{O/H})< -0.08$, i.e., $2\sigma_{Z}$ lower metallicity than expected as the ALM spaxels.

\begin{figure}
\centering
\includegraphics[width=80mm,angle=0]{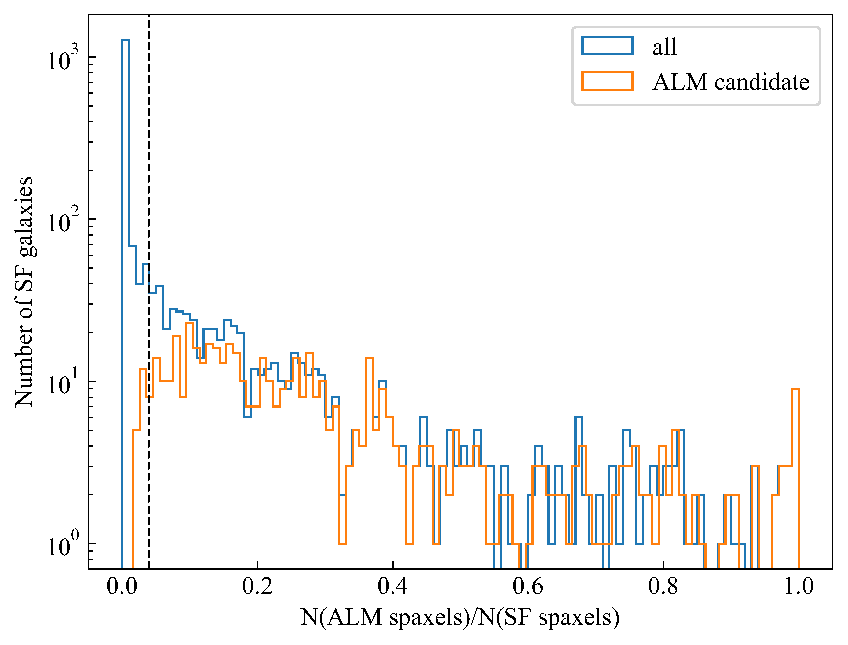}
\caption{Histograms of the ratio of the number of ALM spaxels to the number of SF spaxels in each SFG. The blue histogram represents all SFGs, while the orange one indicates those SFGs having more than 20 ALM spaxels. The dashed line located at 0.04 is the fraction threshold used as the second criterion to define galaxies having ALM regions.}\label{fig3}
\end{figure}

In this work, we focus on SFGs with ALM regions. As the PSF size corresponds to $\sim 20$ spaxels in MaNGA galaxies, ALM cadidate galaxies must contain at least 20 ALM spaxels. In Figure~\ref{fig3}, we show the distribution for the fraction of star-forming spaxels in a galaxy that are classified as ALM. As can be seen, for the majority of SFGs, the ALM spaxel fraction peaks at 0, indicating that they do not contain ALM spaxels. We use a same criterion as \citet{Hwang 2019} to require ALM galaxies to have ALM spaxel fraction $>4$\%, as indicated by the dashed line in Figure~\ref{fig3}. We finally collect a sample of 510 ALM disk SFGs, which is around 1.7 times that of the ALM sample size of \citet{Hwang 2019}. The ALM galaxy fraction is 24\% in our disk SFG sample, which is also in good agreement with that reported in \citet{Hwang 2019}. In the following sections, we refer the non-ALM galaxies as normal SFGs.

We have also tested whether our results are dependent on the applied ALM spaxel fraction threshold in the sample selection. In the test, we applied a stringent ALM spaxel fraction threshold of $>20$\% to select ALM galaxies, yielding a significantly smaller ALM galaxy sample. With this new sample, we perform a similar analysis, finding that the results are largely unchanged.

\subsection{Control Sample} \label{subsec:control}

In the following section, we will compare the properties of ALM galaxies with those of normal SFGs. In order to remove the effects of stellar mass in the comparison, mass-matched control samples are constructed from the normal SFGs.

For an ALM galaxy with a stellar mass of $M_0$, we randomly select a galaxy from the mass-matched normal SFGs to build the control sample. The stellar mass $M_*$ of the matched sample is required to be $\left | \log M_* - \log M_0 \right | < 0.1\, \mathrm{dex}$.  In this way, we create a control sample from normal SFGs, which has a same stellar mass distribution and sample size as the ALM galaxy sample. We also ensure that every matched normal SFG is unique in the final control sample to avoid duplication. This process is then repeated 10 times to construct 10 control samples, which allows a statistically robust comparison in the following analysis.

\begin{figure}
\centering
\includegraphics[width=80mm,angle=0]{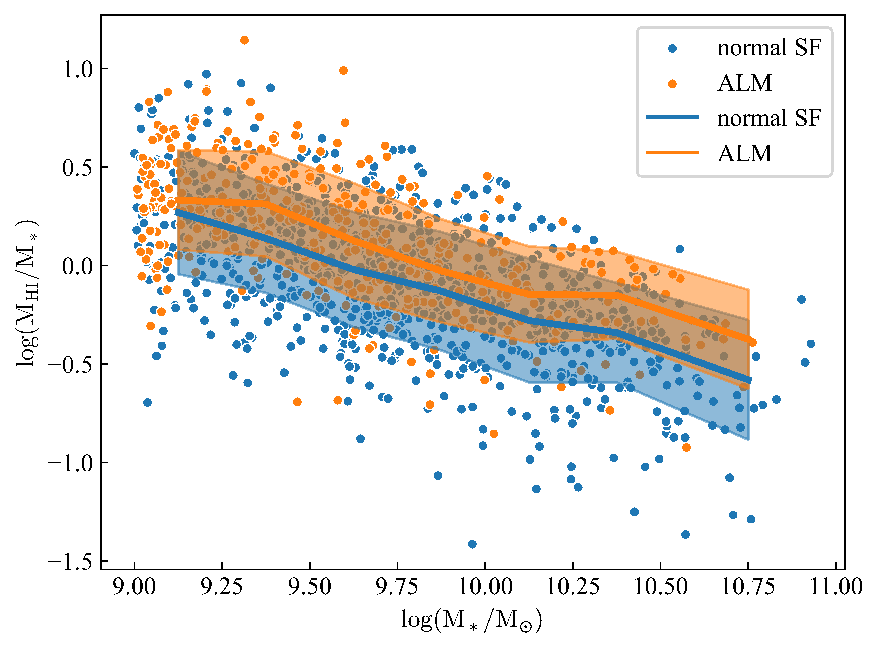}
\caption{The relation between stellar mass and \Hi~gas fraction log($M_{\mathrm{HI}} / M_*$) for ALM galaxies and normal SFGs. ALM galaxies and normal SFGs are shown in orange and blue symbols, respectively. For each sample, the running median value is shown in solid line, and the shaded region indicates the 16-84 percentile.}\label{fig4}
\end{figure}

\begin{figure}
\centering
\includegraphics[width=80mm,angle=0]{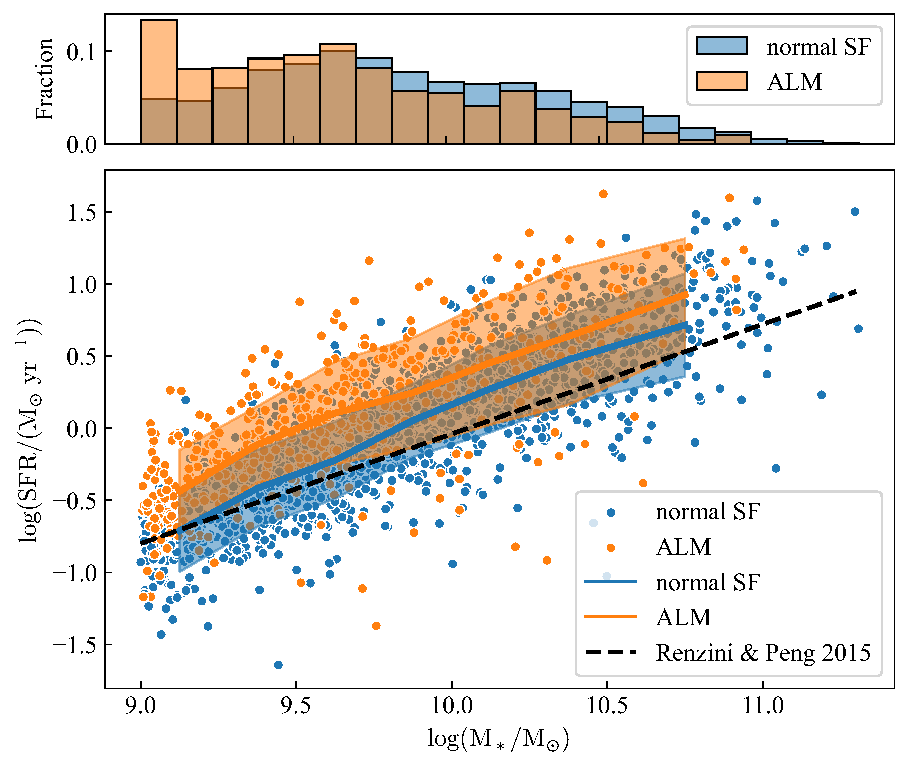}
\caption{Relation between SFR and $M_*$. ALM galaxies and normal SFGs are shown in orange and blue symbols, respectively. For each sample, the running median value is shown in solid line, and the shaded region indicates the 16-84 percentile.}\label{fig5}
\end{figure}

\section{Results} \label{sec:results}
\subsection{\Hi~Content and star formation rate} \label{subsec:hi}

\begin{figure*}
\centering
\includegraphics[width=160mm,angle=0]{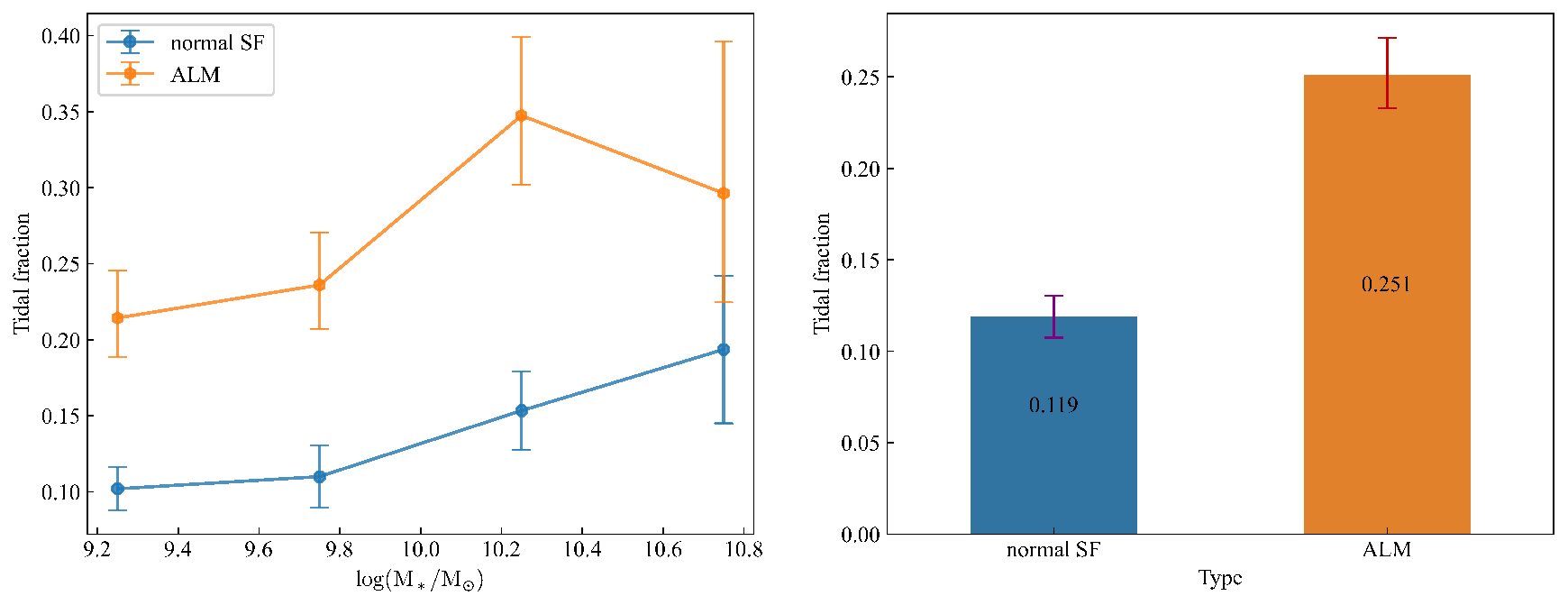}
\caption{Left: tidal fraction as a function of stellar mass. ALM and normal SFGs are shown in orange and blue, respectively. Right: the average tidal fractions for ALM and mormal SFGs. The errorbar is the 1$\sigma$ confidence interval calculated for a binomial population \citep{Cameron 2011}.}\label{fig6}
\end{figure*}

In this section, we firstly investigate the \Hi~gas content of ALM galaxies. At $M_{*}>10^{9}M_{\odot}$, 1306 out of our 2128 disk SFGs have \Hi~detections after cross-matching with the \Hi-MaNGA catalog. In Figure~\ref{fig4}, we show 365 ALM galaxies along with 941 normal SFGs in the
\Hi~gas fraction $M_{\mathrm{HI}} / M_*$ versus stellar mass $M_{*}$ plane. As can be seen, at the stellar mass range probed, ALM galaxies are systematically more rich in \Hi~gas content than normal SFGs. We divide the sample in 7 mass bins within the mass range of log$(M_{\ast}/M_{\odot})=[9.0, 10.5]$ and calculate the enhancement of \Hi~gas fraction of ALM galaxies in each bin. We find that, on average, ALM galaxies have $\sim 0.15$ dex higher \Hi~gas fraction compared to normal SFGs at fixed $M_{\ast}$.

Figure~\ref{fig4} provides key evidence supporting the picture that metal-poor gas inflow results in ALM galaxies. Although the studies of \citet{Hwang 2019} and \citet{Luo 2021} both support this scenario, they do not investigate the cold gas properties of ALM galaxies directly. We also note that some galaxies locate in the most \Hi~rich regime but seem to have normal gas phase metallicity. Such sources are interesting but beyond the scope of this work.

In Figure~\ref{fig5}, we show the SFR$-M_{*}$ diagram for our selected disk SFGs. It is clear that ALM galaxies also have higher SFRs ($\sim 0.24$ dex higher) than normal SFGs at fixed $M_{*}$. This is consistent with previous findings that galaxies with enhanced SFRs also have enhanced cold gas content \citep{Saintonge 2016,Saintonge 2022}. The enhancement of SFR in ALM galaxies is likely related to the recent gas accretion events, although SFR is not directly fed by \Hi~gas.

We note that the selected normal disk SFG sample have slightly higher SFRs compared to the star formation main sequence at $z\sim 0$ \citep{Renzini 2015}, as shown in Figure~\ref{fig5}. This is because we have excluded the bulge-dominated SFGs from analysis, which typically occupy the low-sSFR region at fixed stellar mass \citep{Cano 2019}. One can also see that the stellar mass distribution of ALM sample is biased towards the low mass when compared with the normal SFGs. In observations, low-mass SFGs have lower metallicity than the massive ones \citep{Tremonti 2004,Hidalgo 2017}. Thus a direct comparison between ALM and normal SFG samples is misleading because these two samples have different stellar mass distributions. In the following analysis we use the mass-matched normal SFG control samples for comparison, which can effectively remove the mass selection effect on our results.

\begin{figure*}
\centering
\includegraphics[width=160mm,angle=0]{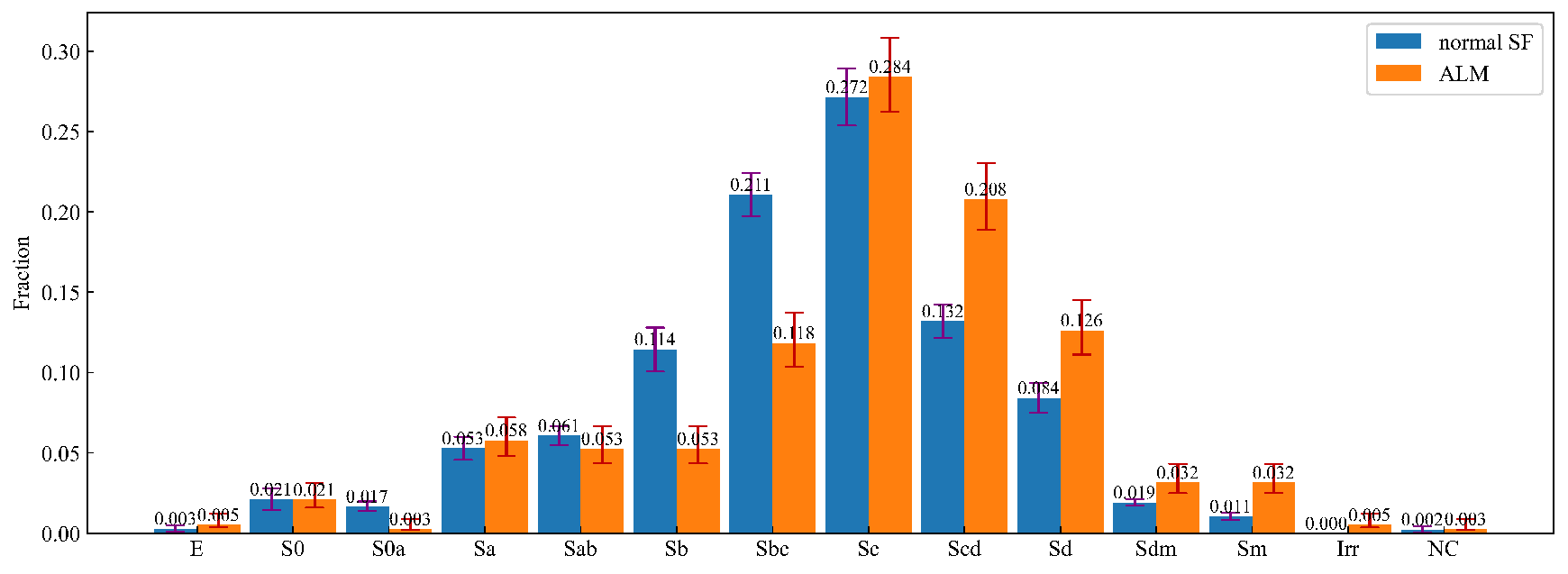}
\caption{Bar plots showing the morphological type distributions of non-tidal normal SFGs and non-tidal ALM galaxies. The height of the blue thick bar indicates the mean value of the type fraction for control sample selected from non-tidal normal SF galaxies and the height of the orange thick bar is the type fraction for non-tidal ALM galaxies. The error bar in purple represents the standard deviation of the fraction for control sample, while the error bar in red indicates the $1\sigma$ confidence interval calculated for a binomial population \citep{Cameron 2011}.}\label{fig7}
\end{figure*}

\begin{figure}
\centering
\includegraphics[width=80mm,angle=0]{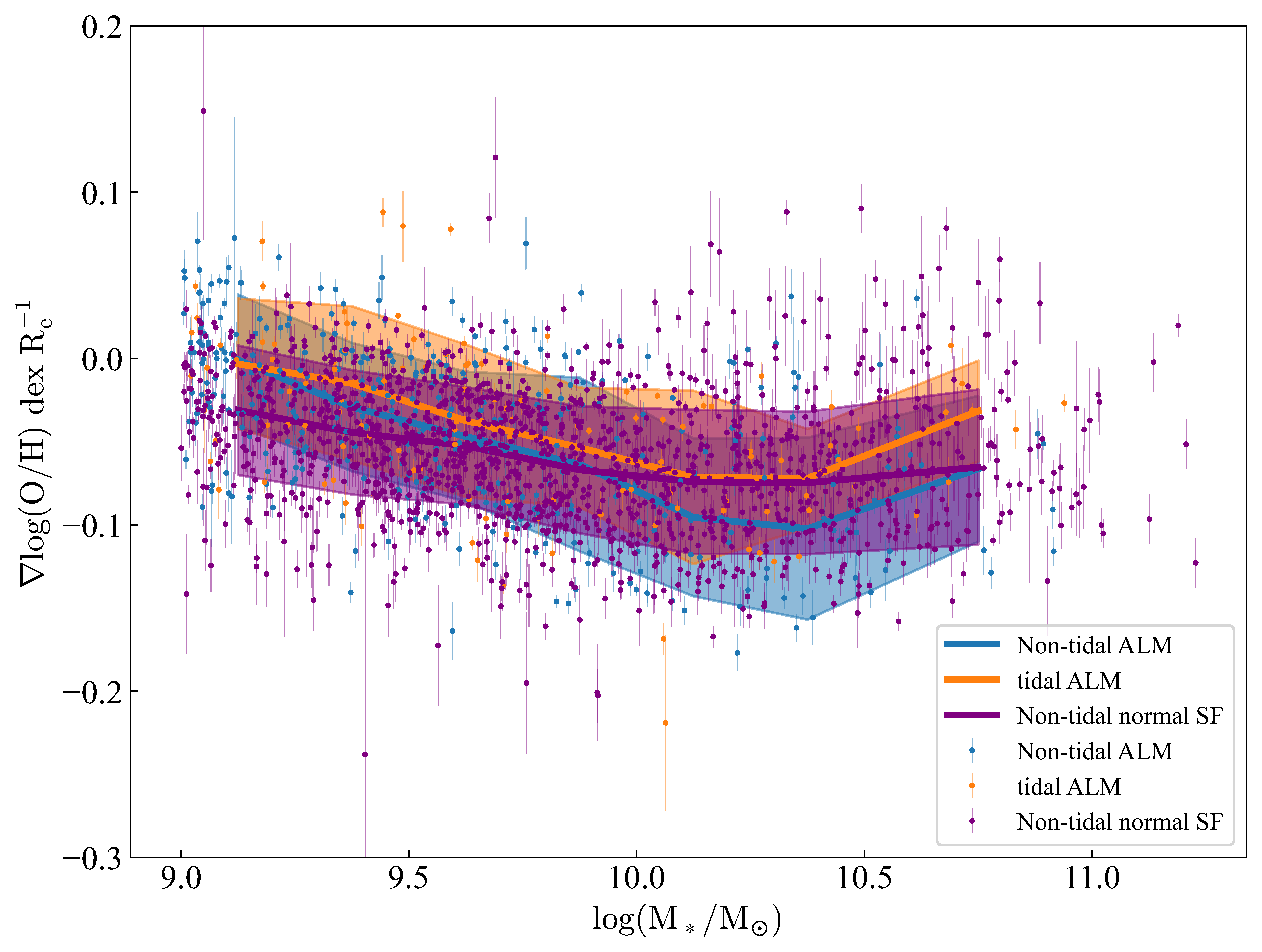}
\caption{The gas phase metallicity gradient as a function of stellar mass. Non-tidal ALM, tidal ALM and non-tidal normal SFGs are shown in blue, orange and purple symbols, respectively. The shaded region indicates the 16-84 percentile for each sample. The running medians of the samples are shown in solid lines.}\label{fig8}
\end{figure}

\subsection{Tidal Fraction} \label{subsec:tidal}
 The instance rate of ALM regions is high in morphological disturbed galaxies, suggesting that galaxy interaction plays a role in the formation of ALM regions \citep{Hwang 2019}. In \citet{Hwang 2019}, morphological disturbed galaxies are identified based on the asymmetry parameter (see their figure 8). As tidal features of galaxies are resulted from galaxy interactions \citep{Yoon 2024}, in this work we compare the tidal fractions of ALM galaxies with those of mass-matched control samples, to access the role of galaxy interaction in the formation of ALM galaxies.

We cross-match the ALM sample with the morphological catalog, yielding 508 galaxies. Then the tidal fractions are calculated in 4 stellar mass bins within the mass range of log$(M_{\ast}/M_{\odot})=[9.0, 11.0]$. The results are presented in Figure~\ref{fig6}. In the left panel, we show tidal fraction as a function of stellar mass. We also derive the tidal fractions for 10 control samples. The average tidal fractions of control samples are shown in blue symbols. It is clear that ALM galaxies have systematically higher tidal fraction. Specifically, the tidal fraction is $\sim 25$\% and $\sim 12$\% in ALM and normal SFGs, as shown in the right panel of Figure~\ref{fig6}. In the  mass range of log($M_{*}/M_{\odot}$)=[10.0,10.5], the tidal fraction of ALM galaxies even reaches $\sim 35$\%.

The high tidal fraction of ALM galaxies suggests that galaxy interaction indeed contributes to the formation of ALM galaxies, as also suggested in \citet{Hwang 2019}. However, the overall tidal fraction is $\sim25$\%, meaning that the majority of ALM systems may not encounter galaxy interactions.  In isolated galaxies, ALM regions are preferentially located in large galactic radius, suggesting that the ALM regions are formed through slowly CGM/IGM accretion \citep{Hwang 2019}. In the next sections, we will focus on non-tidal ALM galaxies.

\subsection{Morphologies and metallicity gradients} \label{subsec:morpho}

Our result suggests that galaxy interaction indeed plays a role in the formation of ALM galaxies. In a merger-free case, ALM galaxies are likely formed through gas accretion from the CGM/IGM. In such a scenario, the plenty cold gas accretion must have left imprints on the morphologies and gas phase metallicity gradients of galaxies. In this section, we will compare the morphologies and gas phase metallicity gradients of non-tidal ALM galaxies with those of normal SFGs.

The MaNGA Visual Morphology catalog have provided the T-Type classification for each galaxy, which is an expansion of the Hubble type \citep{Hubble 1926, Hubble 1927, dev 1977}. Specifically, galaxies are classified into 13 groups, ranging from elliptical (E) to irregular (Irr) types. The catalog also classifies strongly disturbed galaxies of interacting systems as the non-classified (NC) class, as such galaxies are impossible to assign a T-Type.

380 out of our non-tidal ALM galaxies have morphological classifications. We construct 10 control samples from non-tidal normal SFGs using the method mentioned above. The morphological distributions of these two samples are shown in Figure~\ref{fig7}. Both samples have a peak distribution at the Sc type. However, compared to normal SFGs, the morphologies of ALM galaxies clearly shift towards later T-Types, i.e., non-tidal ALM galaxies have more prominent disk components.

Figure~\ref{fig8} shows the gas phase metallicity gradients for ALM and normal SFGs. The gas phase metallicity gradients are drawn from the Pipe3D catalog, which is also based on the RS32 metallicity calibrator. In the Pipe3D pipeline, the gradient is derived based on a linear regression, and the fitting is restricted at $0.5-2.0~R_{\rm e}$. As can be seen, normal SFGs exhibit negative metallicity gradient, which is expected under the "inside-out" disk formation framework \citep{Chiappini 2001,Wang 2011,Sanchez 2014,Pan 2015}. As expected, tidal ALM galaxies exhibit flatter metallicity gradients than normal SFGs, since interaction can trigger gas mixing \citep{Pan 2025}. Interestingly, non-tidal ALM galaxies clearly have steepened gas phase gradient than non-tidal normal SFGs at log$(M_{*}/M_{\odot})>10.0$. This is in line with the finding of \cite{Hwang 2019}, who show that the ALM regions of isolated galaxies preferentially locate at large galactic radius. However, at log($M_{*}/M_{\odot})<9.5$, the trend appears to be reversed. The systematic change in gas phase metallcity gradients from low-mass to high-mass have been revealed in recent IFS studies \citep{Sanchez 2014,Belfiore 2017}. Modeling studies suggest that gas flows driven by stellar feedback are of particular importance in flattening the metallicity gradient in the low-mass regime \citep{Belfiore 2019,Sharda 2024}. The observed high SFRs of ALM galaxies seem to fit this picture.

\begin{figure}
\centering
\includegraphics[width=80mm,angle=0]{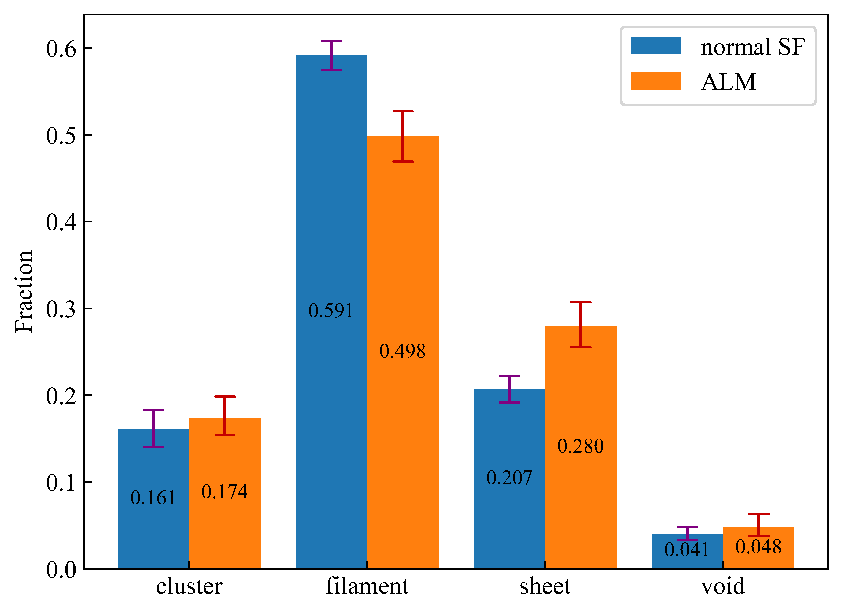}
\caption{Bar plot showing the large-scale structure environment distribution for non-tidal normal SF galaxies and non-tidal ALM galaxies. The LSS environment is categorized into 4 groups: cluster, filament, sheet and void. For each sample, the number marked within the bar shows the fraction of galaxies reside in that LSS environment. The error bar in purple indicates the standard deviation of the fractions calculated from 10 control samples, while the error bar in red indicates the $1\sigma$ confidence interval of the fraction for non-tidal ALM galaxies when treated as beta distribution.}\label{fig9}
\end{figure}

In the current picture of galaxy formation, galaxy interaction is an important channel for building up the bulge of galaxies, while CGM/IGM accretion provides high-angular momentum material that are essential to build up the disk components \citep{Grand 2017}. Figure~\ref{fig7} and ~\ref{fig8} support the CGM/IGM accretion scenario in the formation of non-tidal ALM galaxies, especially for the massive ones.

\begin{figure}
\centering
\includegraphics[width=80mm,angle=0]{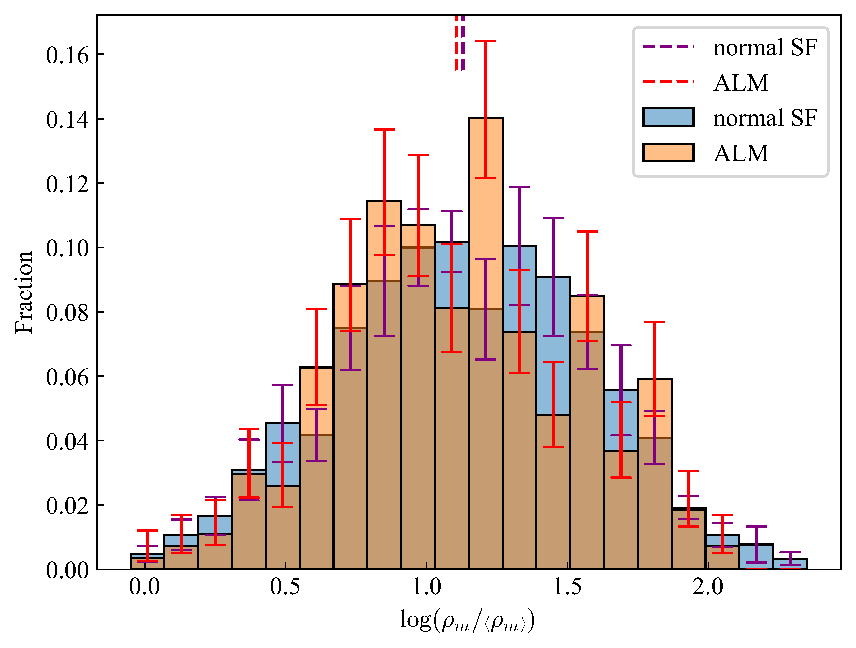}
\caption{Local matter density distributions of non-tidal normal SF galaxies (translucent blue) and the distribution of non-tidal ALM galaxies (translucent orange). Each distribution is normalized so that the sum of the heights of each category equals to 1. The error bar in purple represents the standard deviation of the fraction for control sample in each bin, while the error bar in red indicates the $1\sigma$ confidence interval of the fraction for non-tidal ALM galaxies in each bin when treated as beta distribution. The dashed lines indicate the median values of the two histograms.}\label{fig10}
\end{figure}

\subsection{Environments} \label{subsec:environment}
The environment in which a galaxy reside plays an important role in regulating the gas accretion in that galaxy, as shown in many studies \citep{Stark 2016,Crone 2018,Guo 2020}. In previous studies, the environments of a statistically large sample of ALM galaxies are not explored. In this section, we will investigate both the large-scale structure (LSS) and local density environment of ALM galaxies.

\subsubsection{The Large-Scale Structure environment of ALM galaxies}\label{lss}

The LSS environment classification method used here was from \citet{Hahn 2007}, who used a simple stability criterion from the theory of dynamical systems to distinguish four different LSS: clusters, filaments, sheets and voids. They calculated three eigenvalues of the tidal field tensor $T_1$, $T_2$ and $T_3$ ($T_1 > T_2 > T_3$), where $T_1$ is the eigenvalue of the tidal tensor along the major axis and $T_2$ and $T_3$ are values along the intermediate and minor axes, respectively. These eigenvalues had already been included in the GEMA catalog, with which the LSS classification can be done according to the method introduced by  \citet{Hahn 2007}. More details of the LSS classification can be refereed to \citet{Hahn 2007} and \citet{Wang 2012}.

We cross-match our non-tidal ALM sample with the GEMA catalog, yielding a sample of 293 galaxies. 10 control samples are also constructed from the non-tidal normal SFGs. The LSS environment distributions for the two samples are shown in Figure~\ref{fig9}. As can be seen, for both samples, the majority of galaxies reside in filament and sheet environments. In particular, there is a larger fraction of ALM galaxies reside in sheet environment compared to normal SFGs, although the difference is not very remarkable.

When ranking with matter density, sheets typically have lower matter density than filaments. Figure~\ref{fig9} thus suggests that non-tidal ALM galaxies reside in less-dense LSS environment than normal SFGs.

\subsubsection{Local galaxy environment}
We cross-match our non-tidal ALM galaxies with the Cosmic Slime catalog, yielding a sample of 271 galaxies with local matter density estimates. The local density in the catalog is represented as log($\rho_{m}/\left \langle \rho_{m} \right \rangle$), where $\left \langle \rho_{m} \right \rangle$ is the cosmic mean matter density. We constructed 10 control samples for comparison.

The result is presented in Figure~\ref{fig10}. As can be seen, non-tidal ALM galaxies reside in slightly lower dense regions than normal SFGs. Using a local environment indicator, the result of Figure~\ref{fig10} is in agreement with Figure~\ref{fig9}, showing that the environmental difference between these two samples is most remarkable in the medium-dense environment. We also note that the environmental difference is not very significant, as shown in the median log($\rho_{m}/\left \langle \rho_{m} \right \rangle$) values. In the appendix, we show that the environmental difference appears to be more significant when the O3N2 calibrator is used. To conclude, non-tidal ALM galaxies reside in less-dense local environment than normal SFGs.

\section{Discussion} \label{sec:discussion}
\citet{Hwang 2019} suggest that both galaxy interaction and CGM/IGM accretion contribute to the formation of ALM  galaxies. In this work, we select a large ALM galaxy sample from the completed MaNGA survey, combined with the rich data released in the VAC of SDSS, to further explore the properties of ALM galaxies. In general, our results support the argument of \citet{Hwang 2019} that the ALM regions are the site of newly accreted gas. In this section, we will focus on the new findings of this work.

ALM galaxies have $\sim 0.15$ dex higher \Hi~gas fraction compared to normal SFGs at fixed stellar mass. This is expected under the gas accretion framework. In observations, the \Hi~gas distribution is much more extended than the stellar component \citep{Broeils 1997,Wang 2014,Wang 2016b,Pan 2021}. Without spatially resolved gas map, it is not clear whether the \Hi~gas fraction is also enhanced in the stellar disk of ALM galaxies in which the gas phase metallicity is measured, than in normal SFGs. High resolution maps of gas phase metallicity and cold gas from nearby surveys may provide insights to this issue.

The tidal fraction of ALM galaxies is $\sim 25$\%, while this fraction is only $\sim 12$\% in normal SFGs. This suggests the role of galaxy interaction in the formation of ALM galaxies can not be ignored. Using the asymmetry parameter as a galaxy interaction indicator, \citet{Hwang 2019} found that $\sim 18$\% of their ALM galaxies have strongly disturbed morphologies. Our findings are largely in agreement with \citet{Hwang 2019}. As the identification of tidal feature depends on the quality of the image data, we speculate that not all interacting galaxies are correctly picked out in the morphological catalog. The tidal fraction we derived should be treated as a lower limit of the true one. Nonetheless, the result shown in Figure~\ref{fig6} provides a straightforward view to access the role of galaxy interaction in the formation of ALM galaxies.

The majority of ALM galaxies are thus more likely formed through merger-free processes, such as the CGM/IGM accretion. Our analysis on the morphologies of non-tidal ALM galaxies show that they harbor more late-type morphologies, i.e., they have more prominent disk component. Using the concentration index as a morphological indicator, \citet{Hwang 2019} found that ALM galaxy have similar concentration distribution as the normal SFGs. In \citet{Hwang 2019}, the authors do not remove the morphological disturbed systems from analysis, nor they construct mass-control samples in the comparison. It is thus not surprising that \citet{Hwang 2019} reach a different conclusion.

The environmental analysis shows that non-tidal ALM galaxies reside in lower-dense environment than normal SFGs. This provides the key clue to understand why ALM galaxies exhibit high \Hi~gas fraction. As shown in previous works, galaxies in high density environments (such as galaxy clusters or groups) are often \Hi-deficient, showing asymmetric or even truncated \Hi~disks, which is resulted from gas stripping or strangulation effects in massive, hot halos \citep{Haynes 1984, Koo 2004, Yoon 2017, Zabel 2022, Jim 2023}.  Nevertheless, we find that the environmental difference is not significant between ALM and normal SFGs, both for the LSS and local density environment. This appears in line with some previous studies, which show that environment only plays a mild role in shaping the metallicity distribution of SFGs \citep{Lian 2019,Boardman 2023}. To conclude, our results suggest that environmental condition should have played a role in the formation of ALM galaxies through regulating the ICM/CGM accretion strength.

\section{Conclusions} \label{sec:conclusion}
In this paper, we select a sample of 510 ALM galaxies with $\log(M_* / M_\odot) \ge 9$ from the MaNGA dataset. The \Hi~gas fraction, SFR, tidal fraction, morphology, gas phase metallicity gradient and environment of this sample are explored. Our findings are summarized as follows:

\begin{enumerate}
    \item ALM galaxies have $\sim0.15$ dex higher \Hi~gas fraction than normal SF galaxies at fixed stellar mass. Meanwhile, ALM galaxies have enhanced SFRs ($\sim0.25$ dex higher) than normal SFGs.
    \item  The tidal fraction of ALM galaxies is $\sim 25$\%, twice the value for normal SFGs ($\sim12$\%). This suggests that galaxy interaction is indeed an important factor contributing to the formation of ALM galaxies.
    \item By comparing the morphologies of non-tidal ALM galaxies with those of mass-matched normal SFGs, we find that ALM galaxies exhibit more late-type morphologies, i.e., they have more prominent disk components. At log$(M_{\ast}/M_{\odot})>10.0$, ALM galaxies have steepened gas phase metallicity gradients than normal SFGs. This is consistent with a scenario in which the prominent disk components of ALM galaxies are formed through the slow CGM/IGM accretion from the surrounding environments, at least for the massive disk galaxies.
    \item Compared to normal non-tidal SFGs, there is a larger fraction of non-tidal ALM galaxies reside in sheet-like LSS environment. In terms of local environment, there is also evidence showing that non-tidal ALM galaxies locate in lower dense environment than normal SFGs.
\end{enumerate}

These findings support a picture in which the cold gas accretion from the CGM/IGM plays the major role in the formation of ALM galaxies, while galaxy interaction plays a minor but non-negligible role.

\acknowledgments
We are grateful to the anonymous referee for useful
suggestions that helped improve the presentation of this work. This work is supported by the National Key Research and Development Program of China (2023YFA1608100), the National Natural Science Foundation of China (NSFC, Nos. 12173088, 12233005 and 12073078).

Funding for the Sloan Digital Sky Survey IV has been provided by the Alfred P. Sloan Foundation, the U.S. Department of Energy Office of Science, and the Participating Institutions.SDSS-IV acknowledges support and resources from the Center for High Performance Computing  at the University of Utah. The SDSS website is www.sdss4.org. SDSS-IV is managed by the Astrophysical Research Consortium for the Participating Institutions of the SDSS Collaboration including the Brazilian Participation Group, the Carnegie Institution for Science, Carnegie Mellon University, Center for Astrophysics | Harvard \& Smithsonian, the Chilean Participation Group, the French Participation Group, Instituto de Astrof\'isica de Canarias, The Johns Hopkins University, Kavli Institute for the Physics and Mathematics of the Universe (IPMU) / University of Tokyo, the Korean Participation Group, Lawrence Berkeley National Laboratory, Leibniz Institut f\"ur Astrophysik Potsdam (AIP),  Max-Planck-Institut f\"ur Astronomie (MPIA Heidelberg), Max-Planck-Institut f\"ur Astrophysik (MPA Garching), Max-Planck-Institut f\"ur Extraterrestrische Physik (MPE), National Astronomical Observatories of China, New Mexico State University, New York University, University of Notre Dame, Observat\'ario Nacional / MCTI, The Ohio State University, Pennsylvania State University, Shanghai Astronomical Observatory, United Kingdom Participation Group,Universidad Nacional Aut\'onoma de M\'exico, University of Arizona, University of Colorado Boulder, University of Oxford, University of Portsmouth, University of Utah, University of Virginia, University of Washington, University of Wisconsin, Vanderbilt University, and Yale University.


\software{astropy \citep{astropy 2013,astropy 2018}, matplotlib  \citep{Hunter 2007} }

\appendix

The RS32 metallicity calibrator is used in the main body of this study. In the literature, the O3N2 ((\oiiifull /\hbeta) / (\niifull /\halpha)) calibrator is also frequently used, as it  utilizes the easily accessible and strongest emission lines in the rest-frame optical band and is weakly affected by dust extinction \citep{Marino 2013, Curti 2020}. In this appendix we use the O3N2 calibrator instead to repeat the analysis presented in this work. \citet{Curti 2020} use a polynomial fitting method to calibrate the gas phase metallicity ($Z$=12+log(O/H)) derived from the electronic temperature method and the observed RS32, with the following form
\begin{equation}
    \log(\mathrm{\frac{\text{[O\,{\sc iii}]}\lambda \, 5007}{H\beta} + \frac{\text{[S\,{\sc ii}]}\lambda\lambda \, 6716,6731}{H\alpha}}) = -0.054 - 2.546x - 1.970x^2 + 0.082x^3 + 0.222x^4,
\end{equation}
where $x = Z - 8.69$ and $Z\in [7.6, 8.9]$.

For the O3N2 metallicity calibrator, \citet{Curti 2020} derive a fitting formula with
\begin{equation}
    \log(\mathrm{\frac{\text{[O\,{\sc iii}]}\lambda \, 5007/H\beta}{\text{[N\,{\sc ii}]}\lambda \, 6583/H\alpha}})=0.281-4.765x-2.268x^2
\end{equation}
where $x$ and $Z$ have the same definitions as in equation (1).

\begin{figure}
\centering
\includegraphics[width=80mm,angle=0]{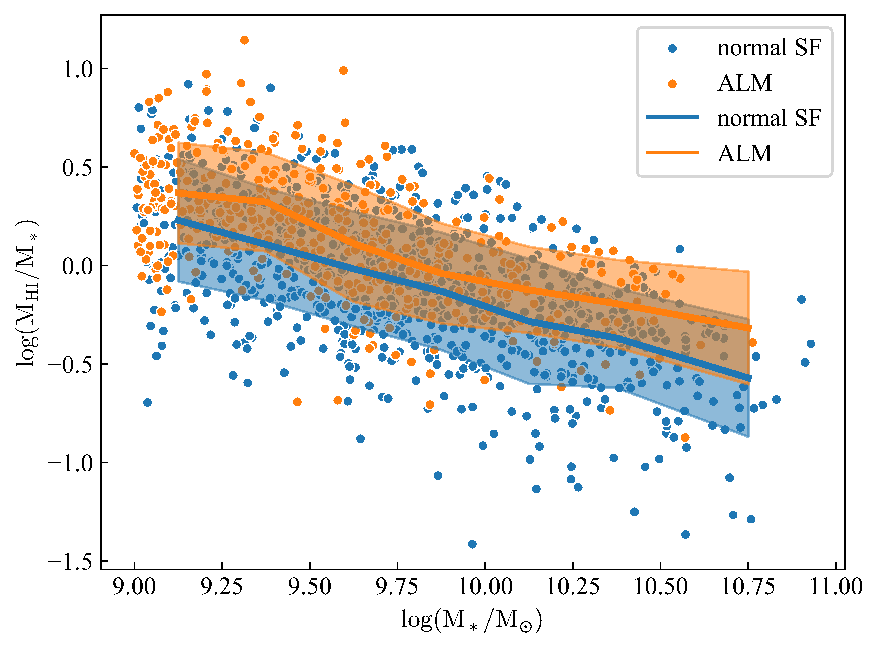}
\caption{Similar to Figure~\ref{fig4}, but with the O3N2 calibrator.}\label{fig11}
\end{figure}

We use a same method to select SFGs and ALM galaxies as mentioned in section 2. For SF spaxels, we have checked and confirmed that the metallicities derived from these two methods are generally in agreement. Therefore, the derived distribution of metallicity deviation is nearly identical as that shown in Figure 2, which also has a $\sigma_{Z}=0.04$ for the normal SF spaxels. With a same ALM galaxy selection method, 584 ALM galaxies are selected by the O3N2 method. Note that the RS32 method yields 510 ALM galaxies. We have checked and found that for the two samples, 475 galaxies are overlapped, implying that the selection of ALM galaxies is not strongly dependent on the applied metallicity calibrator.

Figure 11 to Figure 16 show a similar analysis as presented in the main body of this work, with the O3N2 metallicity calibrator used instead. In general, the results are not changed. For the local density environment distribution, there seems to be a more clear difference between ALM and normal SFGs when the O3N2 calibrator is used. We note that this may be due to the larger ALM galaxy sample selected by the O3N2 method. The O3N2 selected ALM galaxy sample have 323 galaxies with local density estimates, while there is only 271 for the RS32 selected ALM sample. To conclude, the findings of this work are not dependent on the applied metallicity calibrator.

\begin{figure}
\centering
\includegraphics[width=80mm,angle=0]{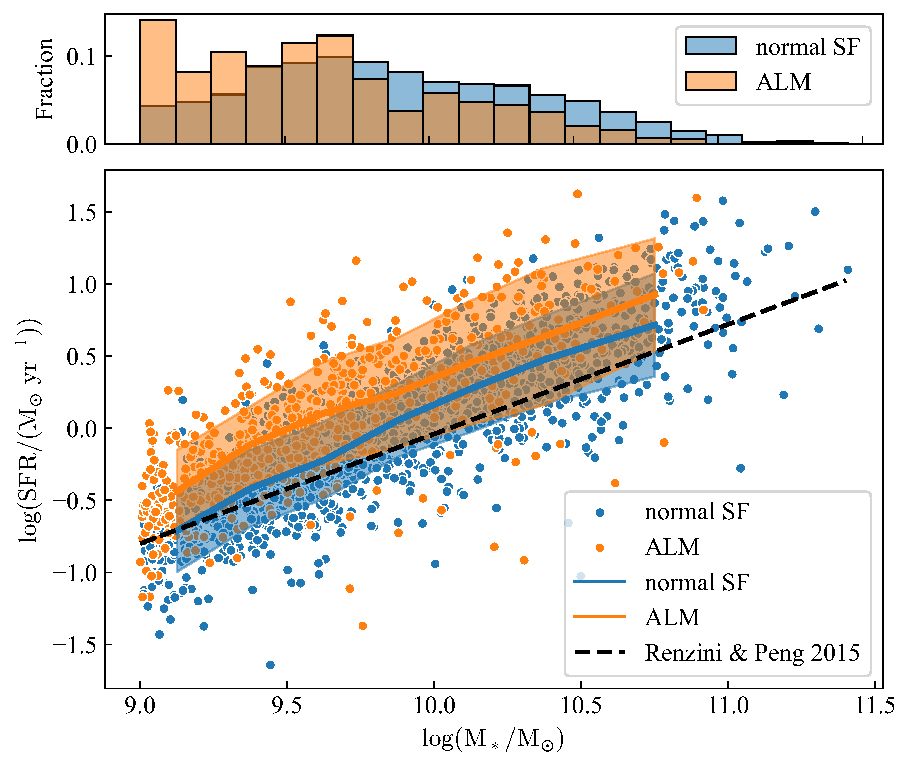}
\caption{Similar to Figure~\ref{fig5}, but with the O3N2 calibrator.}\label{fig12}
\end{figure}

\begin{figure}
\centering
\includegraphics[width=160mm,angle=0]{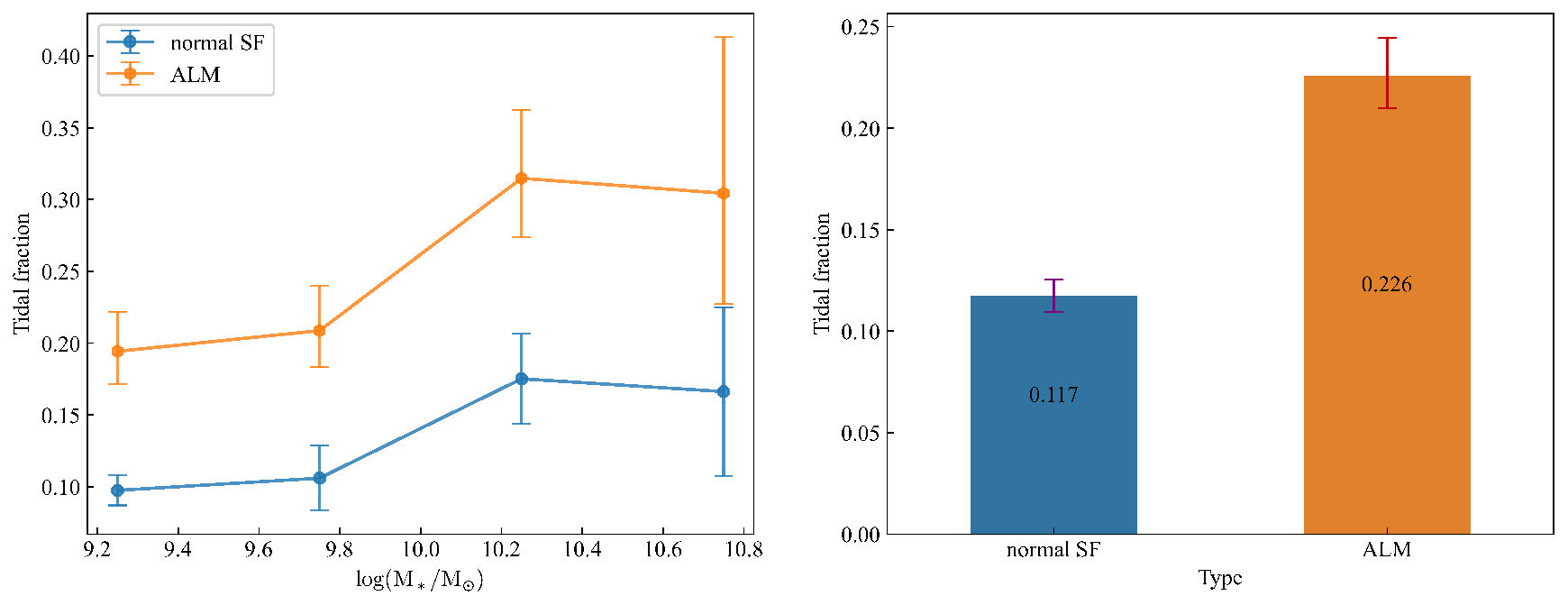}
\caption{Similar to Figure~\ref{fig6}, but with the O3N2 calibrator.}\label{fig13}
\end{figure}

\begin{figure}
\centering
\includegraphics[width=80mm,angle=0]{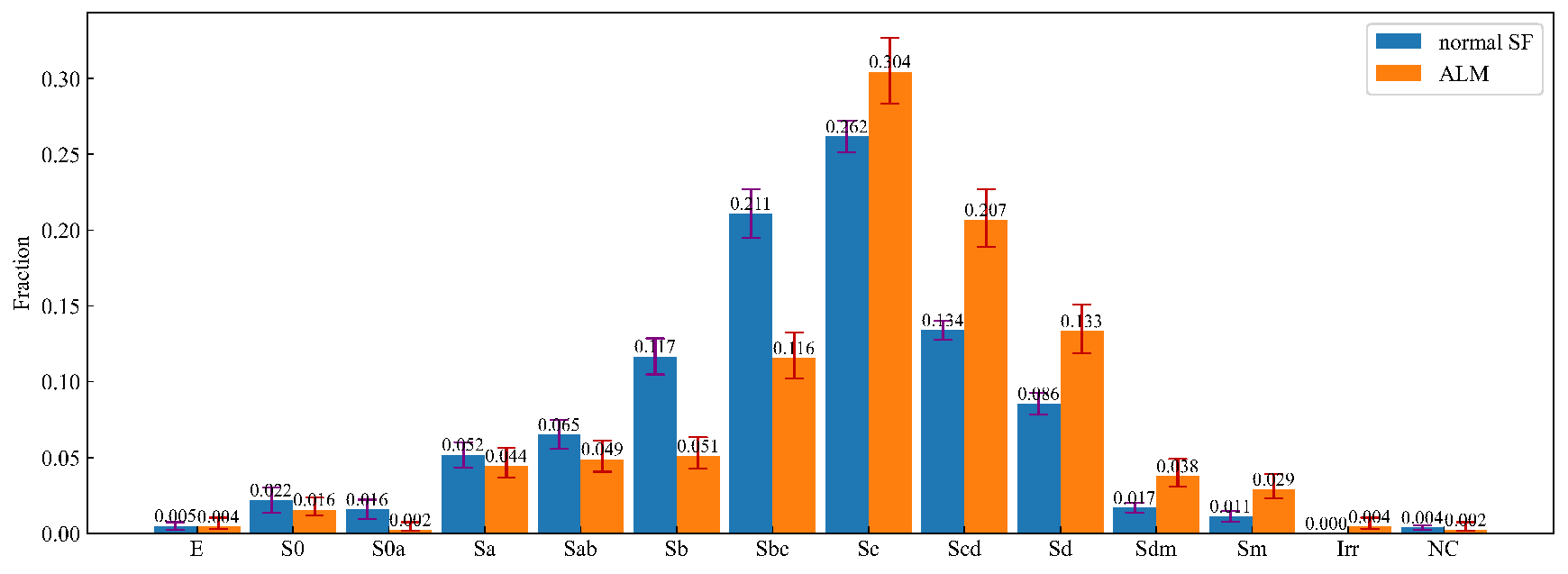}
\caption{Similar to Figure~\ref{fig7}, but with the O3N2 calibrator.}\label{fig14}
\end{figure}

\begin{figure}
\centering
\includegraphics[width=80mm,angle=0]{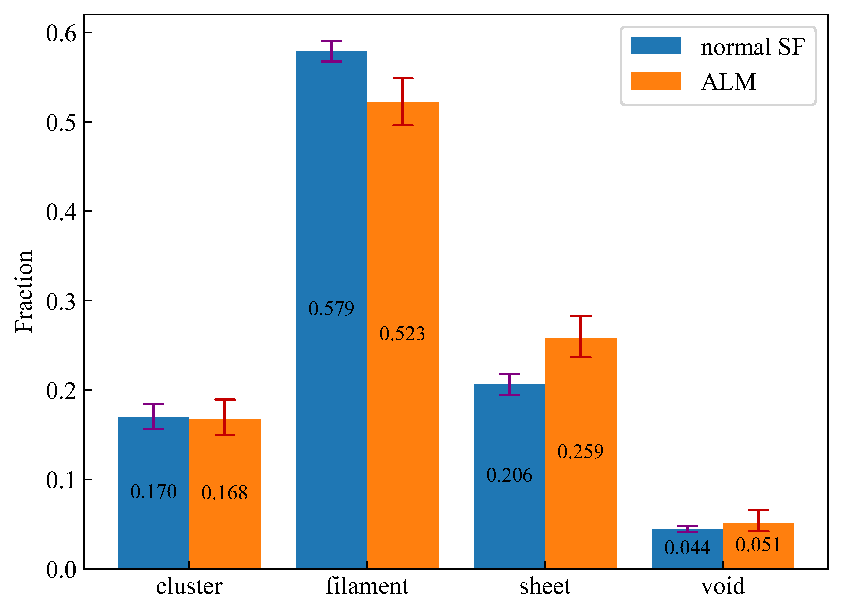}
\caption{Similar to Figure~\ref{fig9}, but with the O3N2 calibrator.}\label{fig15}
\end{figure}

\begin{figure}
\centering
\includegraphics[width=80mm,angle=0]{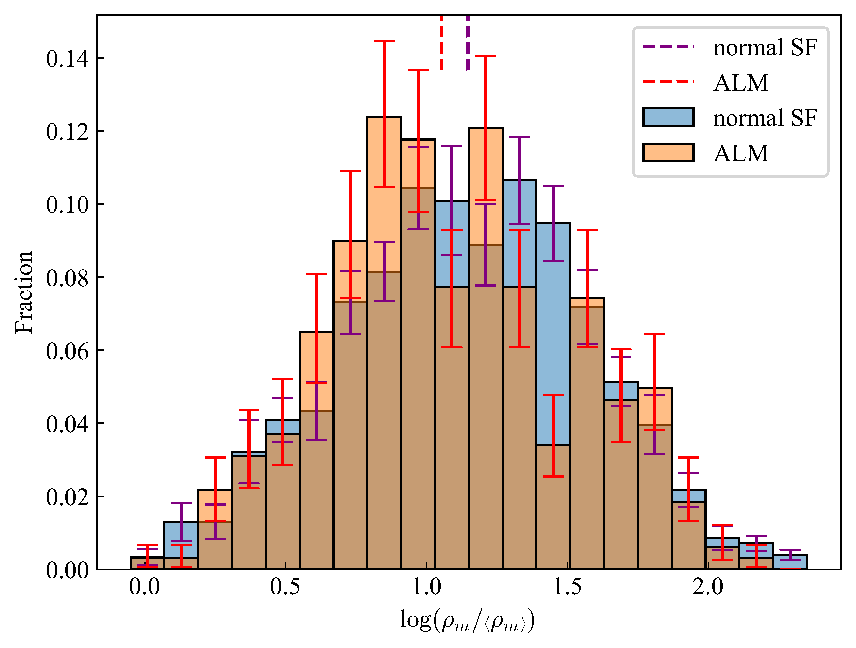}
\caption{Similar to Figure~\ref{fig10}, but with the O3N2 calibrator.}\label{fig16}
\end{figure}


\end{document}